\documentclass{acm_proc_article-sp}
\begin{document}
\setcounter{page}{1}
\pagenumbering{arabic}
\title{A Data Cleansing Method for Clustering Large-scale Transaction Databases\titlenote{This work was supported by the Korea Research Foundation Grant funded by the Korean Government (MOEHRD, Basic Research Promotion Fund) (KRF-2008-331-D00487).}}
\numberofauthors{3}
\author{
\alignauthor
Woong-Kee Loh\\
       \affaddr{Department of Multimedia}\\
       \affaddr{Sungkyul University}\\
       \affaddr{woong@sungkyul.ac.kr}
\alignauthor
Yang-Sae Moon\\
       \affaddr{Department of Computer Science}\\
       \affaddr{Kangwon National University}\\
       \affaddr{ysmoon@kangwon.ac.kr}
\alignauthor
Jun-Gyu Kang\\
       \affaddr{Department of Industrial \& Management Engineering}\\
       \affaddr{Sungkyul University}\\
       \affaddr{jun-gyu.kang@sungkyul.ac.kr}
}
\maketitle
\begin{abstract}
In this paper, we emphasize the need for data cleansing when clustering large-scale transaction databases and propose a new data cleansing method that improves clustering quality and performance. We evaluate our data cleansing method through a series of experiments. As a result, the clustering quality and performance were significantly improved by up to 165\% and 330\%, respectively.
\end{abstract}

\section{Introduction}
\label{sec1}

Data mining has been pursued since the 1990's, and clustering is an important technique in data mining. Clustering is finding the groups of objects having similar features, and it has been rigorously studied~\cite{bex02,hm06,kmss02}, since it has a wide range of applications. Examples of the applications are target marketing and recommendation services. The former is finding groups of customers having similar purchasing patterns and then establishing marketing strategies according to the patterns. The latter is presenting the products to the customer who is highly likely to purchase them according to his/her sales preferences.

Recently, transaction databases have become a new target of clustering~\cite{bex02,grs00}. A {\it transaction\/} is defined as a set of related items, and a {\it transaction database\/} is a database consisting of the transactions obtained in an application~\cite{wxl99,ygy02}. As an example, Figure~\ref{fig01} shows four transactions in a transaction database in the application of search engine services. Each transaction contains the search keywords issued in the same user's session. Another example of transaction database is the product purchase records at a big retail market such as Wal-Mart. In that database, a transaction is defined as a set of products purchased by a customer at a time.

\begin{figure}[t]
\centering
\begin{tabular}{|l|}
\hline
USERID=37264: \\
\hspace{0.2in}amusement park, cherry blossom, mall of america, \\
\hspace{0.2in}entrance fee, disneyland \\
USERID=93272: \\
\hspace{0.2in}freeway, traffic condition, shortcut \\
USERID=20438: \\
\hspace{0.2in}media player, skins, lyric words, download \\
USERID=72620: \\
\hspace{0.2in}major league, ichiro, baseball cap \\
\hline
\end{tabular}
\caption{An example of transaction database.}
\label{fig01}
\end{figure}

Transaction databases have introduced a few technical challenges. First, the objects handled in previous clustering algorithms were represented as $d$-dimensional vectors~\cite{fzs07}. That is, they were represented as the points in $d$-dimensional space and were processed based on the Euclidean distance between them~\cite{grs00}. However, the transactions in transaction databases cannot be represented as $d$-dimensional vectors; they are called {\it categorical data\/}~\cite{hm06}. Second, the size of transaction databases is much larger than the dataset handled in previous algorithms~\cite{nm02}. While the size of dataset in previous algorithms is about several KBs to several MBs, transaction databases have sizes of several GBs up to several TBs.

In this paper, we emphasize the need for data cleansing, which is a pre-processing step before clustering on transaction databases, and propose a new data cleansing method that improves clustering performance and quality. Previous clustering algorithms did not consider data cleansing process. In fact, transaction databases, such as search keyword databases, contain a lot of noise. For example, there are meaningless search keywords such as `tjdnfeorhddnjs' that never appear more than once in the database. This sort of noise causes an increase of the number of useless clusters and the degradation of clustering performance and quality.
A relevant idea is used in information retrieval and text mining. We explain the differences in detail in Section \ref{sec2}.

This paper is organized as follows. In Section~\ref{sec2}, we briefly explain the related work on clustering transaction databases. In Section~\ref{sec3}, we explain the need for data cleansing and propose a new data cleansing method. In Section~\ref{sec4}, we evaluate our data cleansing method through experiments. Finally, we conclude this paper in Section~\ref{sec5}.

\section{Related Work}
\label{sec2}

Most of previous clustering algorithms handled only data objects that can be represented as $d$-dimensional vectors. There are small number of clustering algorithms that handle categorical data or transaction databases, and the most representative one is the ROCK algorithm~\cite{grs00}. It was shown in~\cite{grs00} that we could only get unsatisfactory clustering result on categorical data based on the Euclidean distance. Therefore, ROCK adopted Jaccard coefficient as a similarity measure between categorical data. However, since ROCK has the time complexity higher than $O(n^2)$, where $n$ is the number of objects, it can hardly be applied to large-scale transaction databases.

Efficient clustering algorithms on large-scale transaction data\-bases have been proposed in~\cite{wxl99,ygy02}. A new notion of {\it large item\/} has been proposed in~\cite{wxl99}. For a pre-specified support $\theta (0 \leq \theta \leq 1)$ and a transaction item $e$, if the ratio of clusters containing $e$ in a cluster $C_i$ is larger than $\theta$, the item $e$ is defined as a large item in the cluster $C_i$; otherwise, it is defined as a {\it small item\/}.
The clustering algorithm in~\cite{wxl99}, which we call the {\it LARGE\/} algorithm in this paper, is executed in the direction of maximizing the number of large items and simultaneously minimizing small items by trying to bring the same transaction items together in a cluster.

The CLOPE algorithm~\cite{ygy02}, an improvement of LARGE, is also a heuristic algorithm and maximizes clustering quality by iteration. The algorithm does not use the notion of large/small items; it proposed a more efficient measure for computing clustering quality. CLOPE algorithm was shown in~\cite{ygy02} to have better clustering performance and quality than ROCK and LARGE through a series of experiments.

The problems of LARGE and CLOPE are as follows. The algorithms did not consider the effect of noise data and assumed that the number of result clusters $k$ is very small. However, in actual transaction databases, there contained a lot of noise data with very low frequencies, and the number of result clusters is fairly close to the number of transactions $n$. As a matter of fact, $k$ should be highly variable depending on transactions in the database and items contained in the transactions. If $k$ is very small compared with $n$, the average number of transactions in a cluster should be very high, and such large clusters should have little practical usefulness. LARGE and CLOPE have the time complexity of $O(nk)$, which approaches $O(n^2)$ as $k$ approaches $n$.

In a broad sense, a text database or a document database can be regarded as a form of transaction database; a term and a document correspond to an item and a transaction, respectively. However, these databases have a few essential differences from the transaction databases as the following.

First, since the primary application of text databases is, given a query term, finding and ranking relevant documents, the relevance metrics and feature selection methods are defined between a term and a document~\cite{liu03,tang05}. However, in the transaction database, we use the similarity metrics defined between transactions since we are interested in the relationship between transactions. When clustering documents using relevance metrics such as tf-idf, we should compute the relevance value for each combination of a term and a document, and then we generate feature vectors for each of the documents~\cite{tang05}, which causes severe performance degradation. This preprocessing cost becomes larger when dealing with a larger size of text databases. However, when clustering transactions using inter-transaction similarity metrics, we do not need the preprocessing step of generating feature vectors, and the clustering performance is not severely affected by the size of transaction databases. This advantage of inter-transaction similarity metrics over term-document relevance metrics is more significant when dealing with a frequently updated database. When the database is updated, the entire feature vectors in text database should be re-generated, which is totally unnecessary in the transaction databases.

Second, most transaction databases do not allow duplicated items in a transaction, while any number of same terms can appear in a document in text databases. This causes some relevance metrics useless in transaction databases. For example, for an item $i$ and a transaction $T$, the term frequency is $1/|T|$, where $|T|$ is the cardinality of $T$, i.e., the number of items in $T$, and the inverse document frequency is always identical. Hence, the tf-idf value between $i$ and $T$ is dependent only on the cardinality of $T$; the transaction $T$ of smaller size is regarded to be more relevant to $i$, which is nonsense.

Third, although removing some high frequency and low frequency terms is effective in text databases, the detailed procedure is very different from that in transaction databases. They should be very cautious when removing unnecessary terms in text databases; the terms should not be removed only due to their frequencies, and it is true for both high frequency and low frequency terms. For example, in the world movie database, the term `ponyo' should not be removed only because it appears very rarely, since there should be a lot of people that are interested in the Japanese animation ``Ponyo on the Cliff.'' Removing unnecessary terms in a majority of text databases is controlled under human supervision, which means that it can hardly be fully automated. However, the transaction database has no such issue, and removing unnecessary items can be fully automated. In this paper, we propose a new fully automated data cleansing method with minimal parameter settings and show its effectiveness through experiments.

\section{Data Cleansing}
\label{sec3}

\begin{figure}[t]
\centering
\psfig{file=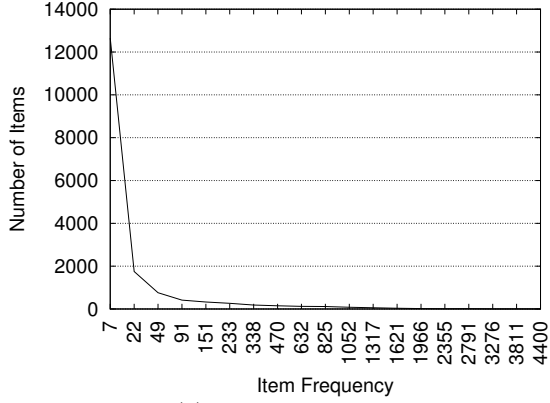,scale=0.31,angle=-90}\\
(a) AOL database.\\[0.1in]
\psfig{file=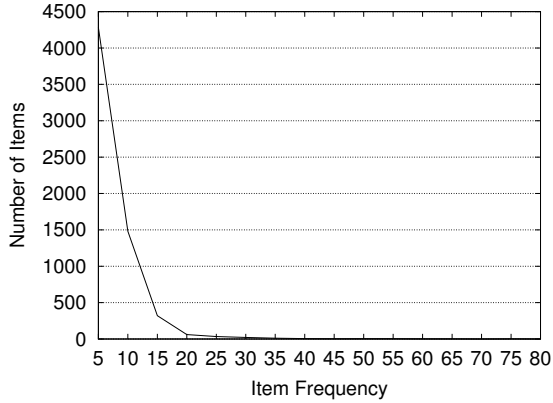,scale=0.31,angle=-90}\\
(b) Keywords database.
\caption{Item frequencies in two real-world transaction databases.}
\label{fig02}
\end{figure}

In this section, we explain the need for data cleansing and propose a new data cleansing method that improves clustering performance and quality. Our data cleansing method decides the usefulness of items according to their frequencies in transactions. Figure~\ref{fig02} shows the item frequencies in two real-world transaction databases. The horizontal axis represents item frequencies, and the vertical axis represents the number of items. As shown in the figure, there exist a lot of items whose frequencies are very small. The two transaction databases are explained in detail in Section~\ref{sec4}.

Transaction items with too low or too high frequencies have negative effects on clustering performance and quality. We explain the phenomenon with examples. We use the same similarity measure between transactions as ROCK as the following Eq.~(\ref{eq1}):
\begin{eqnarray}
\mbox{\it sim\/}(T_1, T_2) = \frac{|T_1 \cap T_2|}{|T_1 \cup T_2|} ~ , \label{eq1}
\end{eqnarray}
where the denominator represents the number of whole items (without duplication) contained in transactions $T_1$ and $T_2$, and the numerator represents the number of items commonly contained in $T_1$ and $T_2$.

First, we explain the effect of the items with too low frequencies. Assume that similarity threshold $\theta$ between transactions is given as $\theta = 0.5$. Consider three transactions $T_1 = \{abcxyz\}$, $T_2 = \{bcdpqr\}$, and $T_3 = \{acdstuvw\}$. Then, for every transaction pair $T_i$ and $T_j$ $(i \neq j, 1 \leq i, j \leq 3)$, it holds that ${\it sim\/}(T_i, T_j) < \theta$, and hence the transactions $T_1$, $T_2$, and $T_3$ does not form a cluster. However, by removing the items with very low frequencies (i.e., $xyzpqrstuvw$), $T_1$, $T_2$, and $T_3$ become $T_1' = \{abc\}$, $T_2' = \{bcd\}$, and $T_3' = \{acd\}$, respectively. Since, for every transaction pair $T_i'$ and $T_j'$, it holds that ${\it sim\/}(T_i', T_j') \geq \theta$, three transactions $T_1'$, $T_2'$, and $T_3'$ should form a useful cluster. In fact, we can easily find enormous number of such transactions as $T_1$, $T_2$, and $T_3$ in real-world transaction databases. The problem due to low frequency items cannot be solved by adjusting or lowering the threshold $\theta$, because the number of low frequency items is not constant across transactions and hence the threshold cannot be fixed.

Second, we show an example where clustering quality is degraded due to the items with too high frequencies. Consider four transactions $T_1 = \{abcdxy\}$, $T_2 = \{cdxyzw\}$, $T_3 = \{qrxyzw\}$, and $T_4 = \{opqrzw\}$. Since, for every transaction pair $T_i$ and $T_{i+1}$ $(1 \leq i < 4)$, it holds that ${\it sim\/}(T_i', T_j') \geq \theta$, it is highly likely that the transactions $T_1$, $T_2$, $T_3$, and $T_4$ should form a large useless cluster $C_L = \{T_1, T_2, T_3, T_4\}$. However, by removing the items with very high frequencies (i.e., $xyzw$), $T_1$, $T_2$, $T_3$, and $T_4$ become $T_1' = \{abcd\}$, $T_2' = \{cd\}$, $T_3' = \{qr\}$, and $T_4' = \{opqr\}$, respectively. The transactions $T_1'$, $T_2'$, $T_3'$, and $T_4'$ naturally form two useful clusters $C_1 = \{T_1', T_2'\}$ and $C_2 = \{T_3', T_4'\}$. Similarly to low frequency items, there are enormous number of transactions such as $T_1$, $T_2$, $T_3$, and $T_4$ in real-world transaction databases, and the problem due to high frequency items cannot be solved by adjusting or raising the threshold $\theta$.

We assume that the item frequency shown in Figure~\ref{fig02} should follow the lognormal or the exponential distribution~\cite{tri01}. Based on this assumption, our data cleansing method performs as the following. First, in the transaction database, we count the number of transaction items for each item frequency (a positive integer value). Next, using the (item frequency, count) pairs, we estimate the parameters such as mean $\mu$ and standard deviation $\sigma$ for the lognormal or the exponential distribution. Finally, for a pre-specified parameter $s$, we remove all the items whose frequencies are either less than $(\mu - s\sigma)$ or greater than $(\mu + s\sigma)$. After removing such items, we also remove empty transactions whose items have been entirely removed. In most cases, $s$ should be 3 $\sim$ 5.

In the case of lognormal distribution, the estimates for two parameters $\mu$ and $\sigma$ are obtained using the following Eq.~(\ref{eq2}):
\begin{eqnarray}
\hat{\mu} = \frac{\Sigma_{i=1..n} \ln x_i}{n} ~ , ~ \hat{\sigma}^2 = \frac{\Sigma_{i=1..n} \left ( \ln x_i - \hat{\mu} \right )^2}{n} ~ , \label{eq2}
\end{eqnarray}
where $n$ is the number of transaction items, and $x_i$ represents item frequency. If there are $k$ items whose frequencies are $x_i$, then $x_i$ appears $k$ times in Eq.~(\ref{eq2}).

In the case of exponential distribution, we compute the estimates for two parameters $\mu$ and $\sigma$ using the following Eq.~(\ref{eq3}):
\begin{eqnarray}
\hat{\mu} = \frac{1}{\hat{\lambda}} ~ , ~ \hat{\sigma}^2 = \frac{1}{\hat{\lambda}^2} ~, \label{eq3}
\end{eqnarray}
where the estimate $\hat{\lambda}$ is computed as the following:
\begin{eqnarray}
\hat{\lambda} = \frac{1}{\bar{x}} ~ , ~ \bar{x} = \frac{1}{n} \Sigma_{i=1..n} x_i ~ . \label{eq4}
\end{eqnarray}

Choosing which of two distributions for a specific transaction database is highly dependent on human expert's view. In our experiments, while choosing any of two distributions contributed to the improvement of clustering quality and performance, the lognormal distribution was more effective. Moreover, improper selection of parameter $s$ value could result in worse clustering performance and quality. Larger $s$ values were advantageous for the lognormal distribution, while smaller $s$ values were advantageous for the exponential distribution.

Our data cleansing method can improve the quality of {\it incomplete\/} clustering results. CLOPE cannot always achieve complete clustering; actually, in most cases, its clustering results are incomplete. In such cases, our method helps improve clustering quality as well as clustering performance.

\section{Evaluation}
\label{sec4}

In this section, we evaluate our data cleansing method through a series of experiments. For our evaluation, we implemented CLOPE~\cite{ygy02} and executed it using real-world transaction databases. We compared clustering quality and performance between two cases: case~(1) using our data cleansing method and case~(2) without using it. In case~(1), the target transaction databases are pre-processed by our data cleansing method and then clustered by CLOPE, while, in case~(2), the databases are directly clustered by CLOPE.

As explained in Section~\ref{sec2}, CLOPE is a heuristic algorithm that enhances clustering quality by iteration. The algorithm computes quality measure called {\it profit\/} of the intermediate clustering result at every iteration, and it stops when the profit does not increase any more. In our evaluation, we use the final profit as the clustering quality measure.

CLOPE receives {\it repulsion\/} $r (> 0)$ as an input parameter. Repulsion is a real value for controlling inter-cluster similarity; higher repulsion implies tighter similarity. Repulsion plays the analogous role of threshold $\theta$ parameter given to ROCK and LARGE, and by adjusting repulsion, we can control the number and quality of clusters.

It was justified experimentally in~\cite{ygy02} that, by using the profit as a metric of clustering quality, CLOPE was more effective than the previous algorithms. In the experiment, CLOPE was run on the mushroom dataset which contains human classification information on poisonous and edible mushrooms. CLOPE achieved the accuracy of 100\% for the repulsion $r \geq 3.1$.

We used two datasets for our evaluation: (a)~AOL search query database and (b)~keyword registration database. The AOL database consists of about 20M queries issued by about 650K users from March 1 through May 31, 2006. The database is a list of records, and every record consists of five fields {\it AnonID\/}, {\it Query\/}, {\it QueryTime\/}, {\it ItemRank\/}, and {\it ClickURL\/}. The first three fields AnonID, Query, and QueryTime represent anonymous user ID, search keyword by the user, and timestamp when the query was issued, respectively. The fields ItemRank and ClickURL are optional, and they appear when the user clicked on any item in query result; they represent the rank and URL of the item clicked by the user, respectively. The keyword registration database is a transaction database; each transaction consists of a URL and a list of registered keywords. The same keyword can be registered by multiple URLs. When a query on a certain keyword is issued, the URLs that registered the keyword are shown in the query result.

We transformed AOL database into a transaction database in the form shown in Figure~\ref{fig01} for clustering by CLOPE. Since a record in AOL database has a query at one time, a user's search queries are spread into multiple records, which appear adjacently in the AOL database. The queries by the same user are collected and a record (transaction) is formed in the transaction database.

We used the user-id field (AnonID) when transforming AOL dataset into a transaction database. A transaction in the transaction database shown in Figure~\ref{fig01} contains all the query terms of the same user-id. The query terms of the same user-id are collected into one transaction, and different transactions have different user-ids. Hence, the inter-transaction similarity based on user-id becomes always zero. We believe that the recommender systems should undergo similar procedures.

The settings for our evaluation are as follows. We used a PC equipped with Intel Core2Quad Q9550 2.83GHz CPU, 4GB RAM, and 600GB HDD and implemented programs using GNU C++ 4.1.2 on CentOS Linux 5.4 64bit Edition with Kernel 2.6.18. We set repulsion for CLOPE as $r = 1.5$, which is a largest value permitted by our system. We assumed that the number of transaction items follow the lognormal distribution and set $s = 5$.

\begin{figure}[t]
\centering
\psfig{file=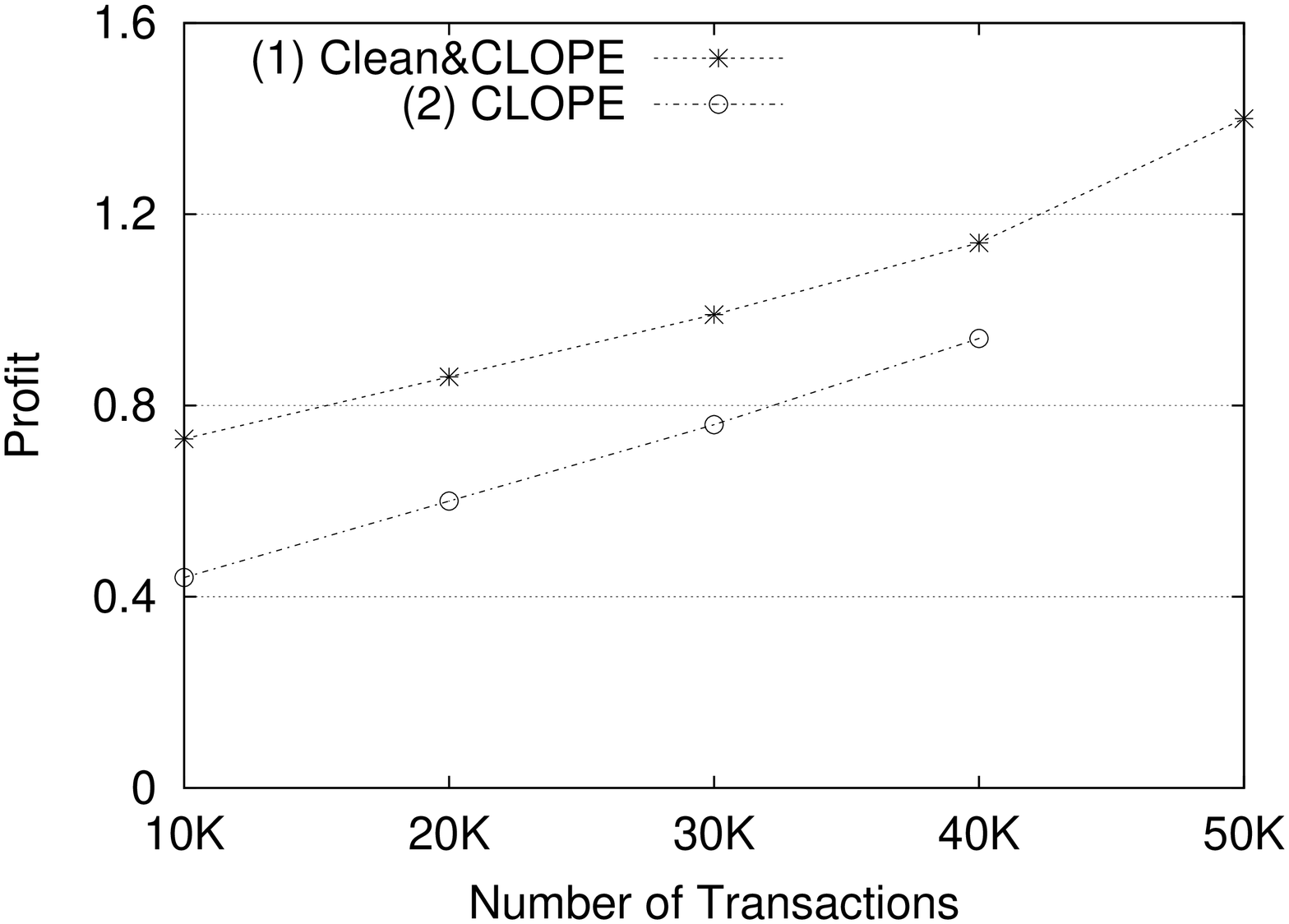,scale=0.31,angle=-90}\\
(a) Clustering quality.\\[0.1in]
\psfig{file=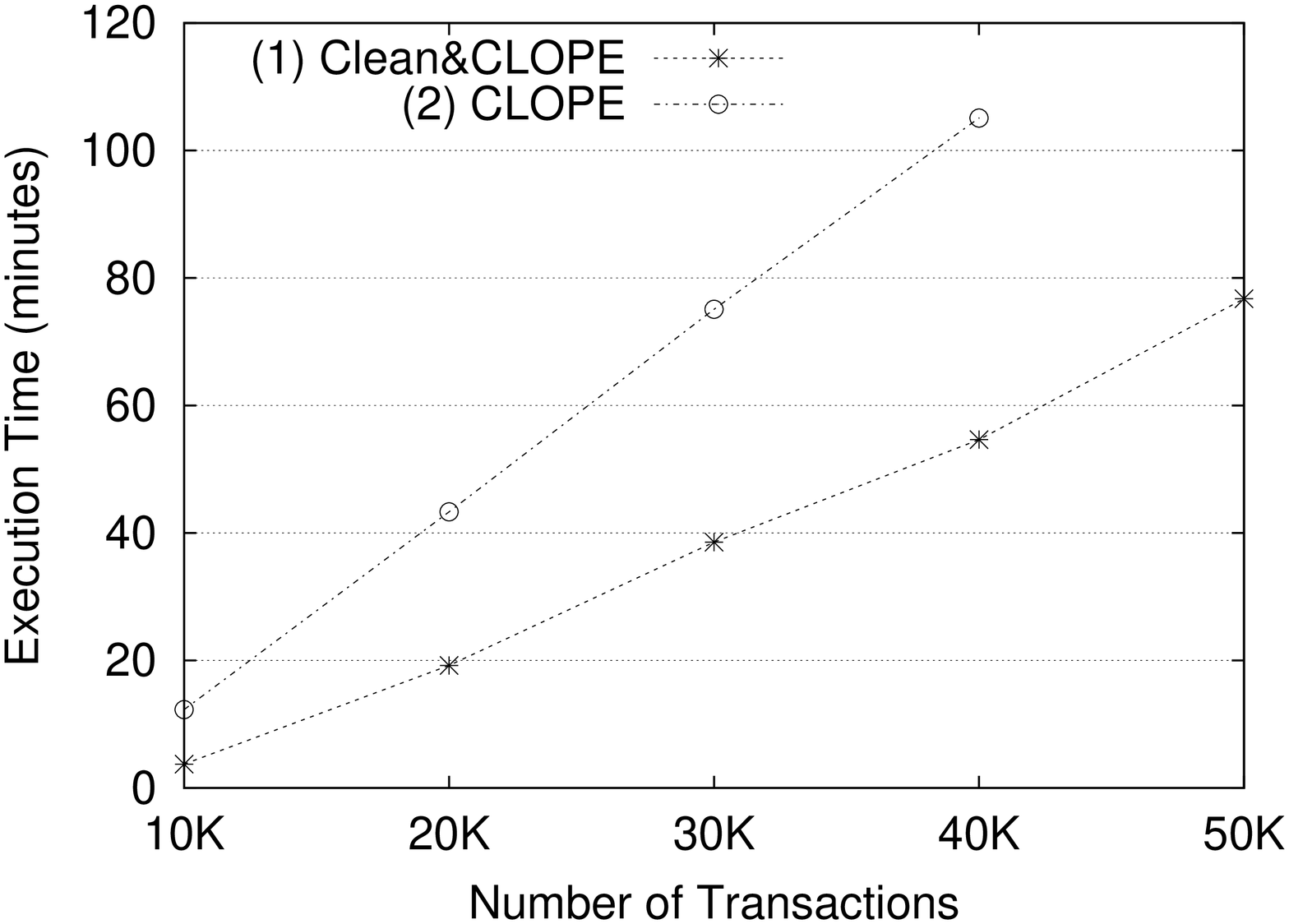,scale=0.31,angle=-90}\\
(b) Clustering performance.
\caption{Comparison of clustering quality and performance using AOL database.}
\label{exp1}
\end{figure}

Figure~\ref{exp1} shows the result of the first experiment using (a)~AOL database; it compares clustering quality and performance between the cases~(1) and (2). In case~(2), for the number of transactions 50K, our program was terminated abnormally, which is most likely due to lack of main memory and swap space.
As shown in the figure, clustering quality and performance was improved by applying our data cleansing method for every number of transactions. The improvement ratio of quality and performance reached up to 165\% and 330\%, respectively. In case~(1), much smaller number $k$ of clusters were formed by CLOPE under the same settings. For that reason, since CLOPE has $O(nk)$ time complexity, we could gain the improvement of clustering performance.

\begin{figure}[t]
\centering
\psfig{file=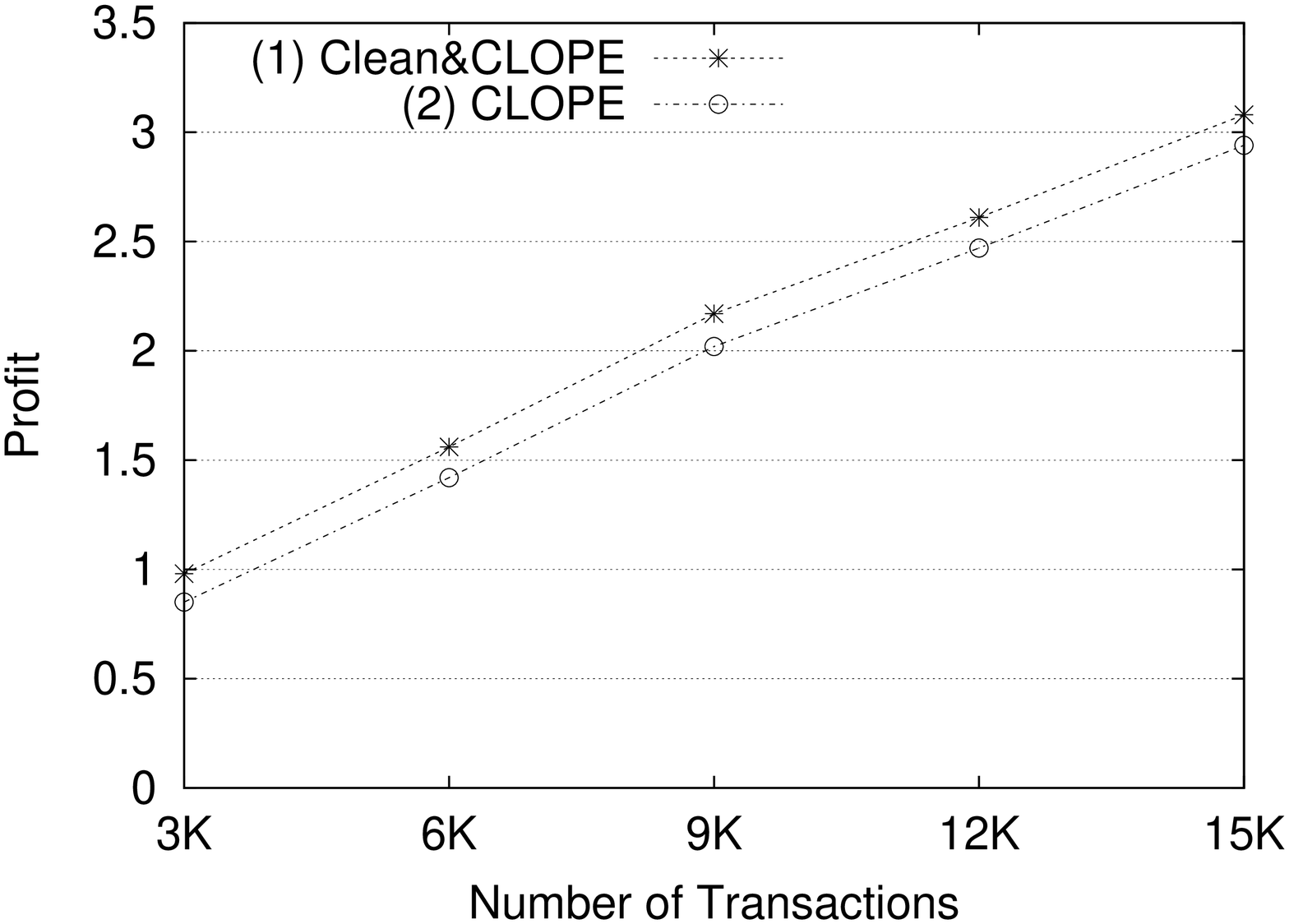,scale=0.31,angle=-90}\\
(a) Clustering quality.\\[0.1in]
\psfig{file=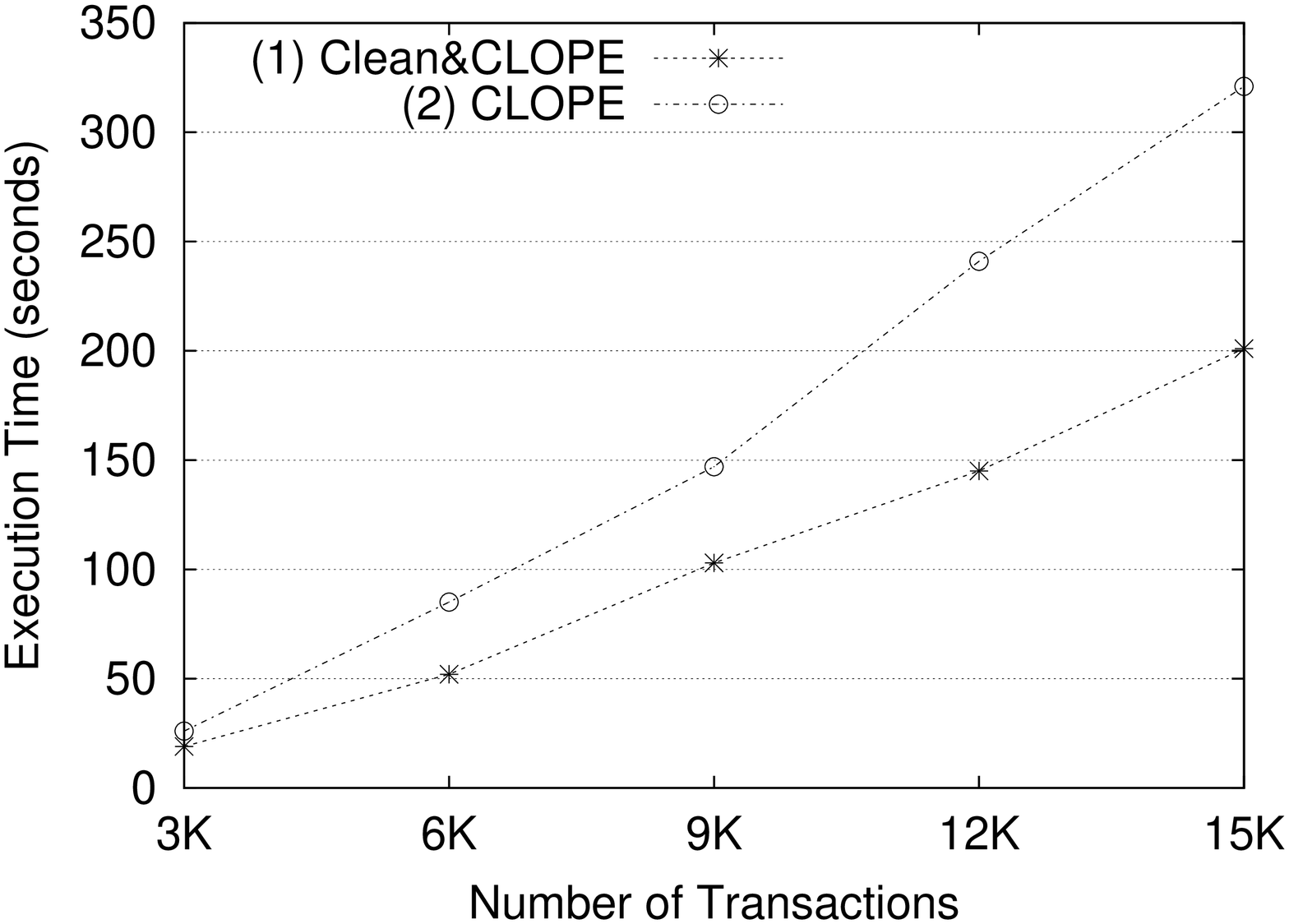,scale=0.31,angle=-90}\\
(b) Clustering performance.
\caption{Comparison of clustering quality and performance using keyword database.}
\label{exp2}
\end{figure}

We performed the second experiment using (b)~keyword database with the same settings as the first experiment, and the result is shown in Figure~\ref{exp2}. As in Figure~\ref{exp1}, clustering quality and performance was also improved by applying our data cleansing method for every number of transactions. The improvement ratio of quality and performance reached up to 115\% and 166\%, respectively.

\begin{figure}[t]
\centering
\psfig{file=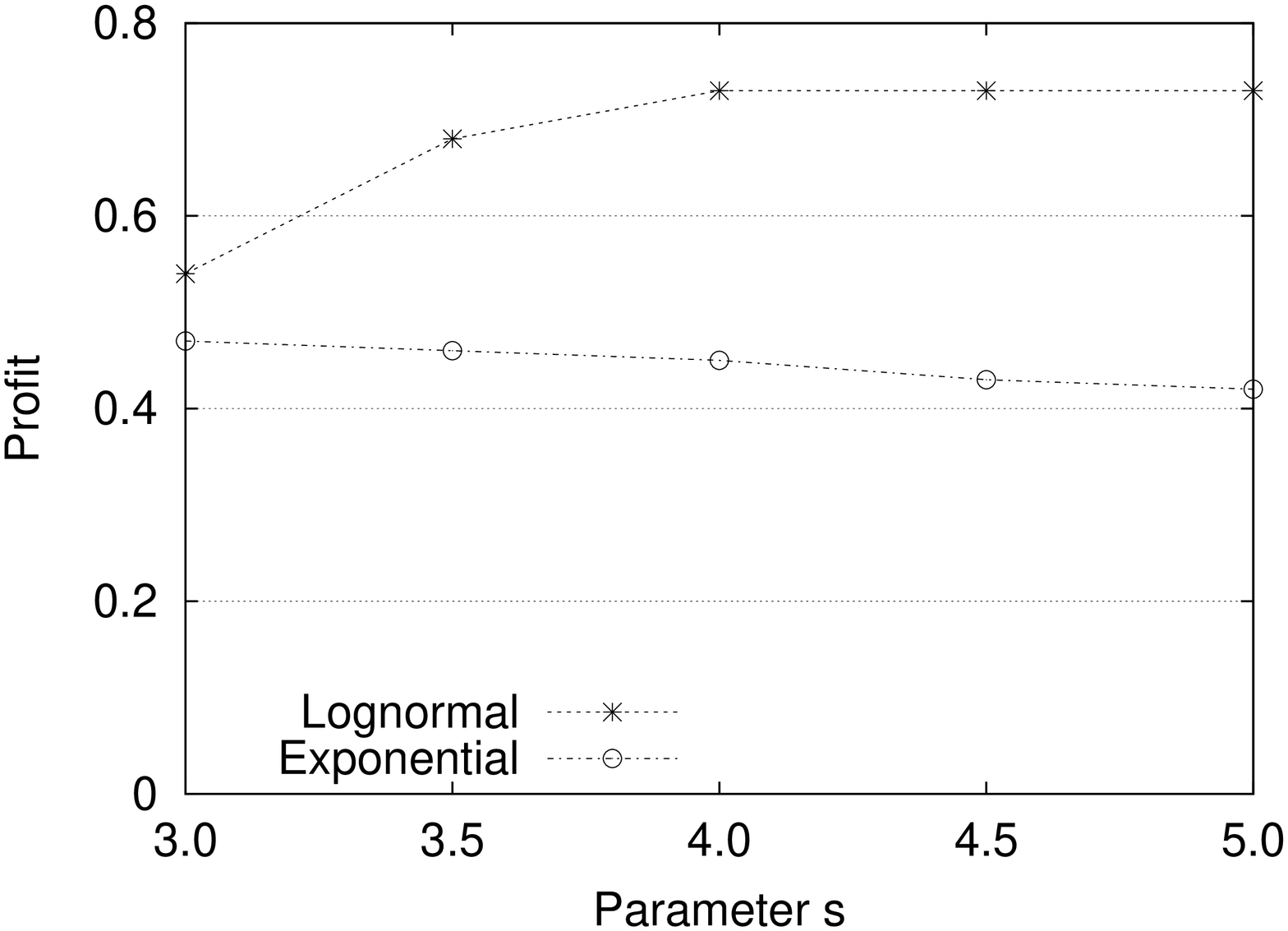,scale=0.31,angle=-90}\\
(a) Clustering quality.\\[0.1in]
\psfig{file=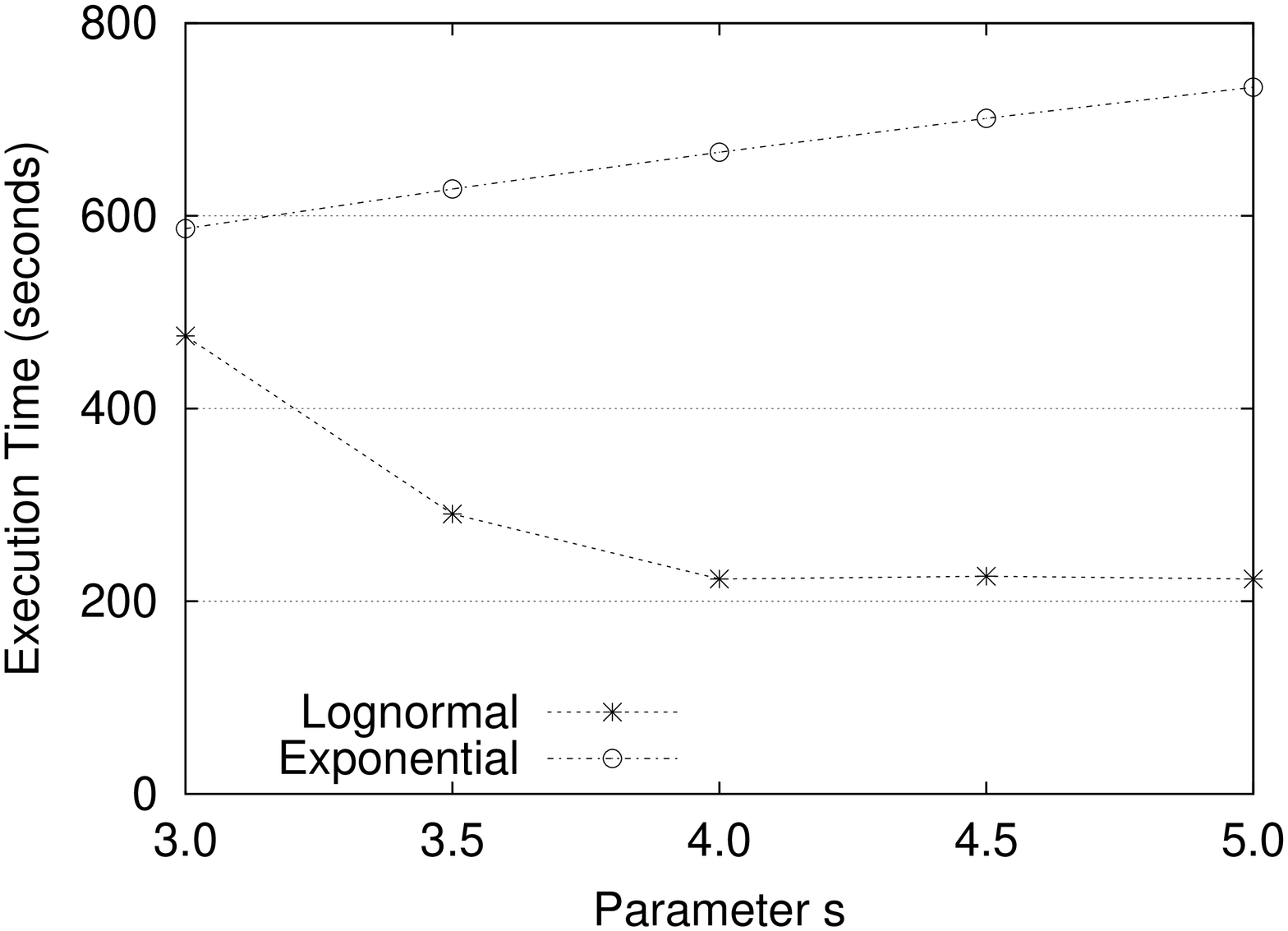,scale=0.31,angle=-90}\\
(b) Clustering performance.
\caption{Comparison of clustering quality and performance for different distributions and $s$ values.}
\label{exp3}
\end{figure}

The third experiment was performed for two distributions and a few parameter $s$ values. We used (a)~AOL database used in the first experiment, and the number of transactions was set as 10K. The experiment result is shown in Figure~\ref{exp3}.
With the lognormal distribution, clustering quality and performance converge to a point for $s$ values larger than or equal to 4.0. This means that there is no improvement in clustering quality and performance by our data cleansing method. With the exponential distribution, smaller $s$ values were advantageous for improving clustering quality and performance.

\section{Conclusions}
\label{sec5}

In this paper, we emphasized the need for data cleansing as a pre-processing step before clustering large-scale transaction databases and proposed a new data cleansing method that improves clustering quality and performance. As the result of our evaluation on our data cleansing method through experiments, the clustering quality and performance were significantly improved by up to 165\% and 330\%, respectively. Although our evaluation was performed by CLOPE, we believe that other clustering algorithms such as ROCK and LARGE should profit by applying our data cleansing method.


\end{document}